\documentclass[%
 reprint,
superscriptaddress,
nobibnotes,
 amsmath,amssymb,
 aps,
prl,
]{revtex4-2}

\usepackage{graphicx}
\usepackage{dcolumn}
\usepackage{bm}
\usepackage{natbib}
\usepackage{multirow}
\usepackage{amssymb} 
\usepackage{amsmath} 
\usepackage{xspace}
\usepackage{float}

\newcommand{\pt}{\ensuremath{p_{\mathrm{T}}}\xspace}
\newcommand{\ptmiss}{\ensuremath{\pt^\text{miss}}\xspace}
\newcommand{\ttbar}{\ensuremath{\text{t}\bar{\text{t}}}\xspace}

\begin{document}

\title{Optimal transport for a novel event description at hadron colliders}

\author{L. Gouskos}
\affiliation{European Organization for Nuclear Research (CERN), Geneva, Switzerland}
\author{F. Iemmi}
\email{fabio.iemmi@cern.ch}
\affiliation{Institute of High Energy Physics (IHEP), Beijing, China}
\author{S. Liechti}
\affiliation{University of Zurich (UZH), Zurich, Switzerland}
\author{B. Maier}
\email{benedikt.maier@cern.ch}
\affiliation{European Organization for Nuclear Research (CERN), Geneva, Switzerland}
\affiliation{Karlsruhe Institute of Technology (KIT), Karlsruhe, Germany}
\author{V. Mikuni}
\affiliation{National Energy Research Scientific Computing Center, Berkeley Lab, Berkeley, USA}
\author{H. Qu}
\affiliation{European Organization for Nuclear Research (CERN), Geneva, Switzerland}



\begin{abstract}
We propose a novel strategy for disentangling proton collisions at hadron colliders such as the LHC that considerably improves over the current state of the art. Employing a metric inspired by optimal transport problems as the cost function of a graph neural network, our algorithm is able to compare two particle collections with different noise levels and learns to flag particles originating from the main interaction amidst products from up to 200 simultaneous pileup collisions. We thereby sidestep the critical task of obtaining a ground truth by labeling particles and avoid arduous human annotation in favor of labels derived \emph{in situ} through a self-supervised process. We demonstrate how our approach—which, unlike competing algorithms, is trivial to implement—improves the resolution in key objects used in precision measurements and searches alike and present large sensitivity gains in searching for exotic Higgs boson decays at the High-Luminosity LHC.
\end{abstract}

\maketitle


\section{Introduction}
\label{sec:intro}

At the High Luminosity-LHC (HL-LHC), up to 200 proton collisions will take place simultaneously. This poses an unprecedented challenge to the reconstruction algorithms of experiments such as ATLAS and CMS, hindering their ability of searching for new physics with highest sensitivity if collision products are not disentangled properly. A performant rejection of particles from subordinate proton collisions (pileup) is therefore paramount to the success of the LHC physics program. 
Exploiting the excellent position resolution of tracking systems, charged particles from pileup collisions can be effectively mitigated~\cite{ATLAS:2015ull,CMS:2017yfk} by discarding  particles not associated to the primary vertex. 
Contributions from neutral particles (photons and neutral hadrons), however, 
can only be reconstructed using the calorimeter systems with comparably poor spatial resolution. To this end, dedicated algorithms based on a physics-motivated, rule-based selection were developed to obtain a set of per-particle probabilities indicating whether neutral particles originate from the leading primary vertex or not. For instance, the \textsc{PUPPI} algorithm is the state-of-the-art at the CMS experiment~\cite{Bertolini:2014bba,CMS:2020ebo}.

The complexity of the pileup mitigation task motivated the proposition of machine learning (ML)-based algorithms: the PUMML algorithm~\cite{Komiske:2017ubm} relies on image recognition techniques to identify particles stemming from pileup vertices; other algorithms, e.g., in Refs.~\cite{ArjonaMartinez:2018eah,Maier:2021ymx}, exploit graph neural networks (GNNs) or transformers. In all cases, the ML-based approaches yield sizable improvements compared to rule-based algorithms.

One of the main limitations of these ML algorithms is that they require a sound definition of a ground truth, i.e., the assignment of a label to each particle, indicating if it originates from the leading primary vertex or not, to train a fully-supervised ML model. Such labels are available in simplified detector simulations, such as the ones implemented in Delphes \cite{delphes}. However, due to merged energy deposits and the much more complex event reconstruction, an unambiguous ground truth definition for neutral particles is intractable in data and in the full-scale simulations based on Geant4 \cite{AGOSTINELLI2003250} that are used by experiments such as ATLAS and CMS. Therefore, per-particle target labels based on human annotation are very hard to obtain and exhibit insufficient sharpness, rendering fully supervised strategies suboptimal. Thus, none of the previously described ML-based solutions can be implemented in a straightforward manner by the experiments at the LHC. Recently, it has been proposed in~\cite{Li:2022omf} to train a semi-supervised network using charged particles, whose provenance can be obtained under the assumption of perfect tracking, and to extrapolate the results to neutral particles. While this approach would allow to train on full-scale simulations and on data, it only works in a central detector region where tracking is available, whereas pileup is predominant in the forward regions of the detectors. 
Therefore, our primary motivation to develop the algorithm for Training Optimal Transport with Attention Learning (\textsc{TOTAL}) in this article is to address these bottlenecks. We achieve this by employing  metrics inspired by optimal transport (OT)~\cite{monge1781memoire} problems as the cost function of a self-supervised, attention-based GNN, whose architecture closely follows the development in~\cite{Mikuni:2020wpr}. This network is used to assimilate two simulated samples, one containing only the particles arising from the primary interaction and the other containing also contributions from pileup. By training in a self-supervised fashion, i.e., not relying on per-particle truth labels from human annotation, our approach can be realistically implemented in full-scale simulations and does not rely on any kind of extrapolation. Following this strategy, we are able to derive an event description that is straightforward to implement and that exhibits greatly improved precision, yielding a global sensitivity enhancement for SM measurements and searches alike, explained as follows.


\section{Sliced Wasserstein distance}
\label{subsec:swd}

Instead of relying on truth labels for reconstructed particles, we design an alternative objective that transforms the entire set of particles from multiple simultaneous interactions into the same collision event containing only the primary interaction. Borrowing from concepts of OT, we are interested in finding the transport function that leaves particles with similar features between sets unchanged while removing contributions from additional interactions. The Wasserstein distance~\cite{villani2008optimal} can leverage geometric information in the probability space to estimate the distance between probability measures.
Given two probability measures $\alpha \in P(X)$ and $\beta\in P(Y)$, both defined in the set of probability measures $P(\Omega)$, we identify the $q$-Wasserstein distance problem as the determination of the transportation plan $\gamma \in \Gamma(\alpha,\beta)$ that satisfies:


\begin{equation}
    \mathrm{OT}_q(\alpha,\beta) = \left ( \underset{\gamma \in \Gamma(\alpha,\beta)}{\mathrm{inf}} \int_{\Omega\times\Omega}c(\mathbf{x,y})^qd\gamma(\mathbf{x,y}) \right )^{1/q},
    \label{eq:OT}
\end{equation}
where $c$ is the cost function evaluated over observations $\mathbf{x}, \mathbf{y}$ drawn from the sets $X, Y$ and the order $q$ can be chosen to give the $q$-th root of the total cost incurred.
In high-dimensional spaces, solving the OT problem becomes computationally expensive since only the one-dimensional case allows a closed-form solution~\cite{wass1}.
This observation motivates the formulation proposed in~\cite{wass1,wass2} to introduce the sliced Wasserstein distance (SWD) as an integral over one-dimensional transport problems:


\begin{equation}
    \mathrm{SWD}(\alpha,\beta) = \int_{\mathcal{S}^{d-1}}\mathrm{OT}_q(\mathcal{R}_{\theta}(\alpha),\mathcal{R}_{\theta}(\beta))d\theta.
\end{equation}
The term $\mathcal{R}_{\theta}$ represents the linear operation that projects (or ``slices") the probability measures over the one-dimensional space and is integrated over a uniform measure $\theta$ in the unit sphere ${\mathcal{S}^{d-1} \in \mathbb{R}^{d}}$. The dimensionality $d$ corresponds to the number of particle features considered. Since the integral is intractable, we can instead use a Monte Carlo approach to replace it with multiple random projections. In this formulation, the optimal coupling is the one that minimizes the cost function evaluated over one-dimensional, sorted projections, thus replacing the expensive OT problem by several (one for each projection), yet simple, sorting problems.



Accordingly, we denote the set of particles in the sample with pileup as ${x_p\in\mathbb{R}^{N\times d}}$, while the same particle collision in the set without pileup is denoted as ${x_{np}\in\mathbb{R}^{N\times d}}$. Since the set of particles without pileup is considerably smaller than the one with, we keep the overall number of particles $N$ fixed by zero-padding as necessary. Given $M$ projections with permutations $\mu$ and $\nu$ that sort $N$ particles of the sets $\mathcal{R}_{\theta_m}(x_{p})$ and  $\mathcal{R}_{\theta_m}(x_{np})$, the SWD is calculated as:

\begin{equation}
    \mathrm{SWD}(x_p,x_{np}) = \frac{1}{M}\sum_j^M\sum_i^N c\left ( \mathcal{R}_{\theta_j}(x_{p,\mu(i)}), \mathcal{R}_{\theta_j}(x_{np,\nu(i)})\right),
    \label{eq:swd}
\end{equation}
where the cost function $c(\mathbf{x,y}) \equiv |\mathbf{x-y}|^2$ is used in the following studies.


Our method aims to train a neural network to output a set of weights $\omega \in [0,1]$, for each particle, that removes particles from pileup collisions.  While the network is trained using a larger set of features, the SWD calculation in Eq.~\ref{eq:swd} is carried out using only the four-vector $(p_x,p_y,p_z,E)$ for each particle. The weights $\omega$ are learned by minimizing $\mathrm{SWD}(x'_p,x_{np})$, with the weighted set $x'_{p} = \omega x_{p}$. By multiplying the four-vector with a single weight we are able to re-scale the magnitude of the momentum vector while preserving the direction in the calculation.

As a result, particles created from pileup collisions increase $\mathrm{SWD}(x'_p,x_{np})$ and should  populate $\omega$ values closer to zero, whereas particles from the primary collision are assigned $\omega\approx 1$.

An extra term can be added to Eq.~\ref{eq:swd} to control the energy scale of the events in the sample containing pileup. To that end, a constraint on the missing transverse momentum (\ptmiss) is introduced to the loss function as the mean square error (MSE) of the \ptmiss values between samples:

\begin{equation}
    \mathcal{L} = \mathrm{SWD}(x'_p,x_{np}) + \lambda\, \mathrm{MSE}({\ptmiss}(x'_p), {\ptmiss}(x_{np})).
    \label{eq:loss}
\end{equation}

The parameter $\lambda$ controls the strength of the regularization. In this letter, we present results for $\lambda =0$ (no regularization), and for $\lambda=10^{-3}$, resulting in the \ptmiss constraint to have the same order of magnitude as the SWD term.


We simulate proton collisions at a center-of-mass energy $\sqrt{\text{s}}=\text{14}$\,TeV employing the Pythia v8.244 event generator~\cite{pythia8,Corke_2011}. The Delphes v3.4.3pre01 detector simulation~\cite{delphes} is used to obtain reconstructed particles with a detector layout resembling the Phase-II upgrade of the CMS detector. The simulated physics processes include jets produced via quantum chromodynamics (QCD); the production of a top quark-antiquark pair (\ttbar), where both W bosons decay leptonically; the vector boson fusion (VBF) production of a Higgs boson decaying into undetectable dark matter particles; the production of a heavy resonance ($\textrm{Z}^\prime$) decaying to \ttbar, where both W bosons decay hadronically; and the production of a W boson in association with at least one jet (W$+$jets). The former three processes are used for training and inference, while the $\textrm{Z}^\prime$ and W$+$jets processes are used to assess the performance on jet substructure and the robustness of the algorithm, respectively.

Starting from the same simulated hard interaction, we generate two event samples. The first one consists solely from particles produced during the hard interaction process. In the second sample, we add contributions from pileup before reconstructing the event. The number of pileup interactions follows a Poisson distribution with a mean of 140 to match HL-LHC conditions. 

The TOTAL algorithm accomplishes the task of pileup mitigation by taking the available properties of each particle as input features. These are the four-vector (${\pt, \eta, \phi, E}$), the impact parameters in the transverse plane and along the beam axis, the particle ID, the electric charge, and the vertex identification for charged particles. The inputs are pre-processed so that the order of magnitude of all features is $1$. We use the output weights $\omega \in [0,1]$ to re-scale the four-vector of each particle in the slicing process. Per event, we consider the first 9000 particles (sorted by descending \pt and including zero-padding) and gather the 20 nearest neighbors in the $\eta$--$\phi$ plane for each particle when building the graph. The SWD in Eq.~\ref{eq:loss} is computed taking 128 random projections per collision event. Different choices of  number of projections were tested and changes in the results were found to be negligible when using more projections. 

\section{Results}
\label{sec:results}

The network trained with $\lambda=0$ (cf. Eq.~\ref{eq:loss}) is evaluated on samples statistically independent from the ones used during training. The per-particle weights obtained from the evaluation process are used to re-scale the four-momenta of the particles in the event. We then use the FastJet software~\cite{Cacciari_2012} to cluster two \textsc{TOTAL} jet collections using the anti-$k_t$ algorithm~\cite{Matteo_Cacciari_2008} with small and large radius parameters of $R=0.4$ and $R=0.8$, respectively. The set of re-scaled particles is also used to compute \ptmiss. We evaluate the network on three benchmark processes: QCD multijet production, which is ubiquitous at hadron colliders and constitutes an important background to many measurements and searches; dileptonic \ttbar, which often enters analyses both as a signal or as a background process; and $\textrm{Z}^\prime$, which is enriched in events with hadronically decaying top quarks merged in large-radius jets and is well suited to study jet substructure. These processes cover different physics scenarios, namely cases where sizable \ptmiss comes from detector and reconstruction inefficiencies, and events where genuine \ptmiss is caused by neutrinos.


\begin{sloppypar} 
To compare the performance of different pileup mitigation algorithms, a matching in ${\Delta \mathrm{R} = \sqrt{(\eta_i - \eta_j)^2 + (\phi_i - \phi_j)^2}}$ is performed between reconstructed jets and generator-level jets. The latter are obtained by clustering only the particles coming from the leading primary vertex, before any detector and reconstruction effects. The following studies are based on generator-matched jets, namely reconstructed jets for which a generator-level jet is found within $\Delta \mathrm{R} < 0.3$.
\end{sloppypar}

\begin{figure*}[htb!]
    \centering
    \includegraphics[width=0.32\textwidth]{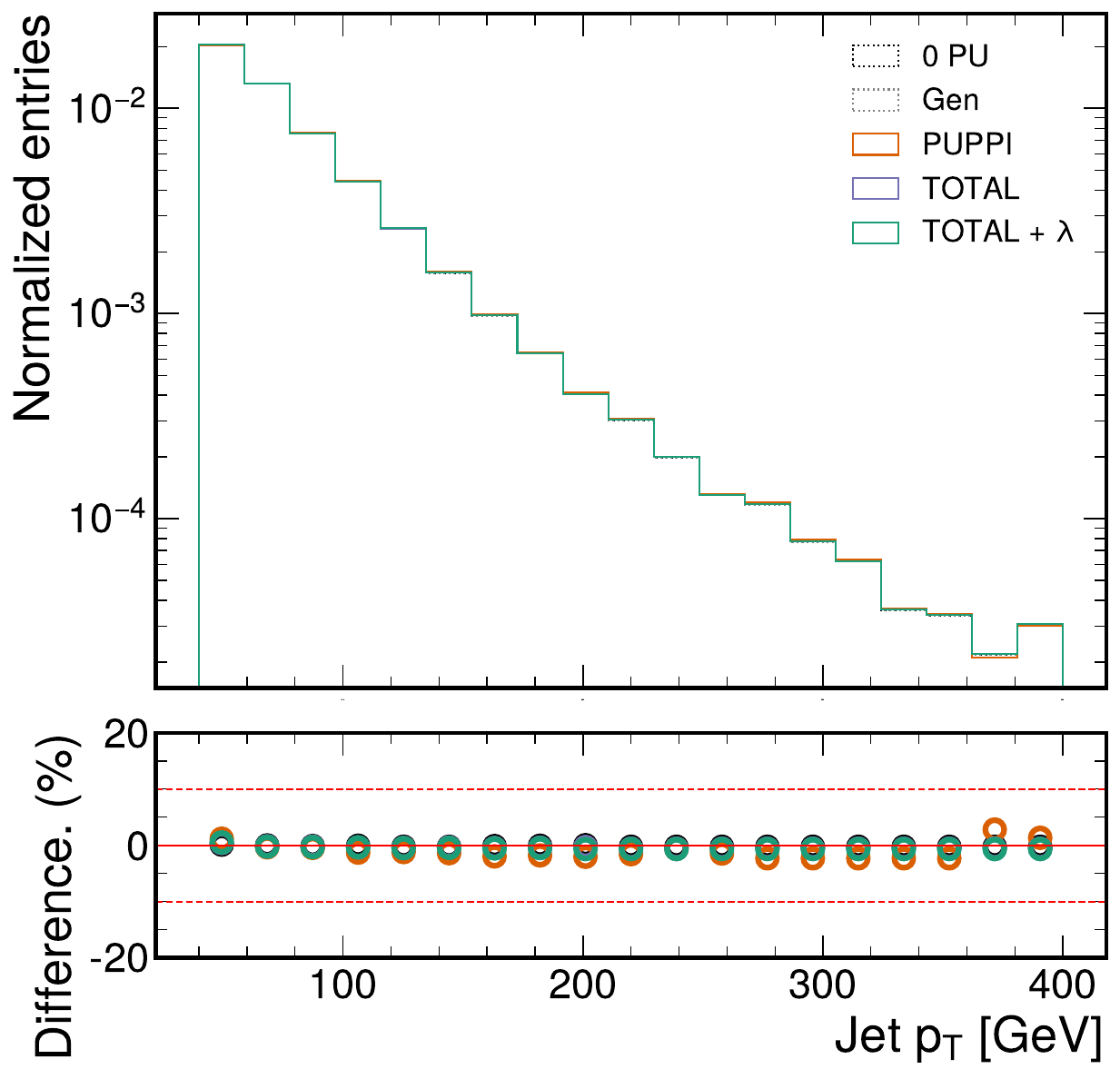}
    \includegraphics[width=0.32\textwidth]{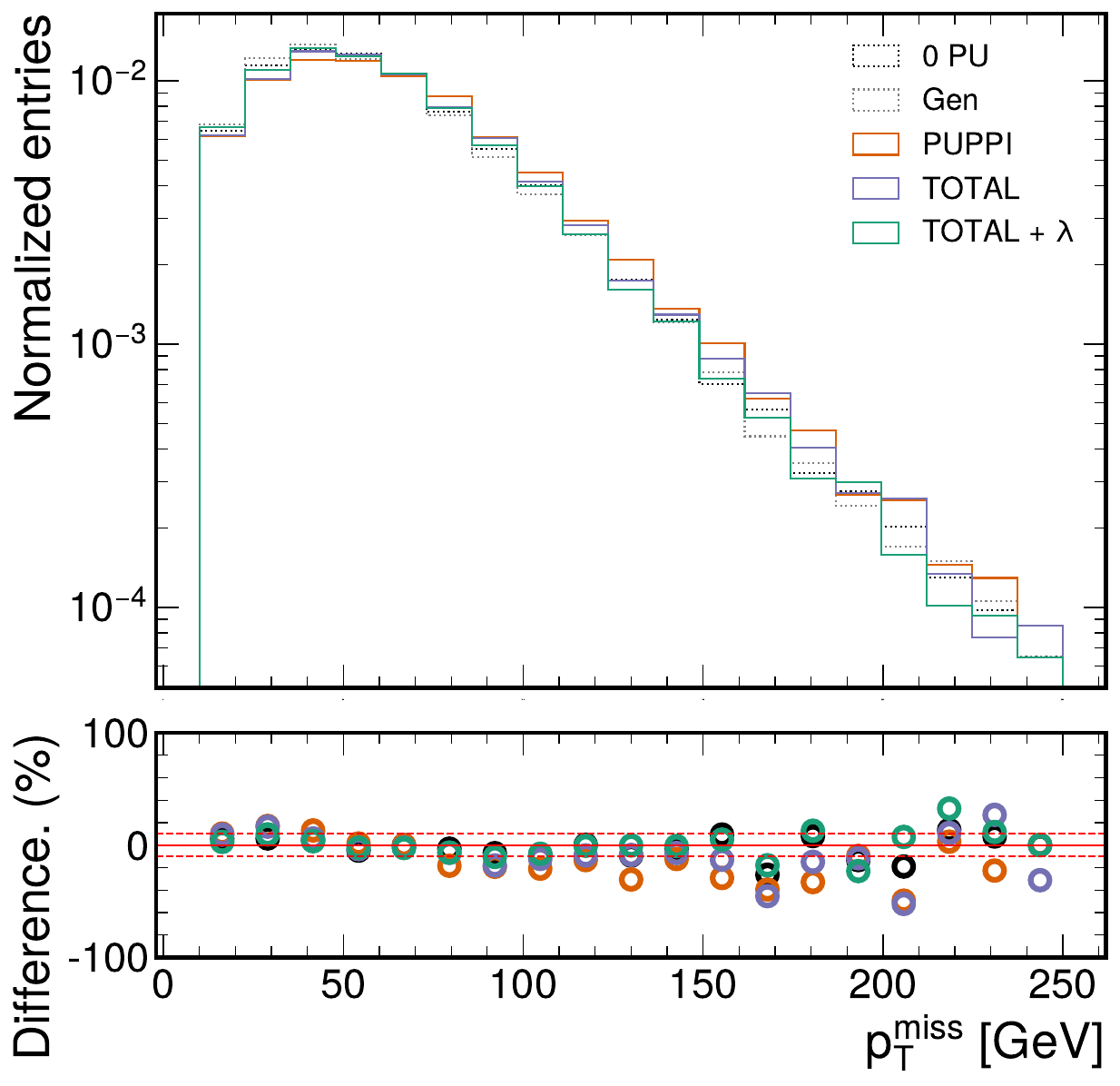}
    \includegraphics[width=0.32\textwidth]{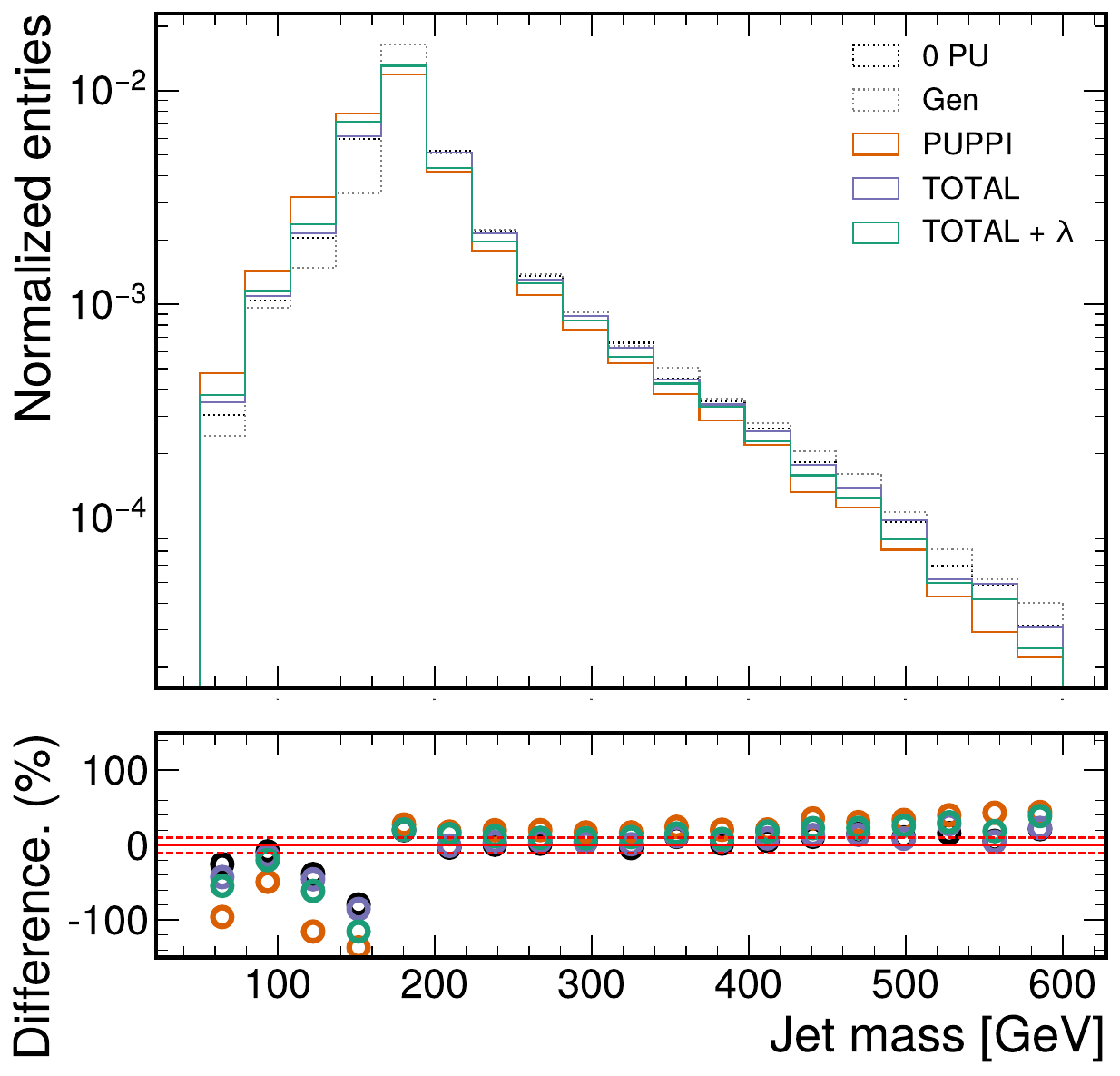}
    \includegraphics[width=0.323\textwidth]{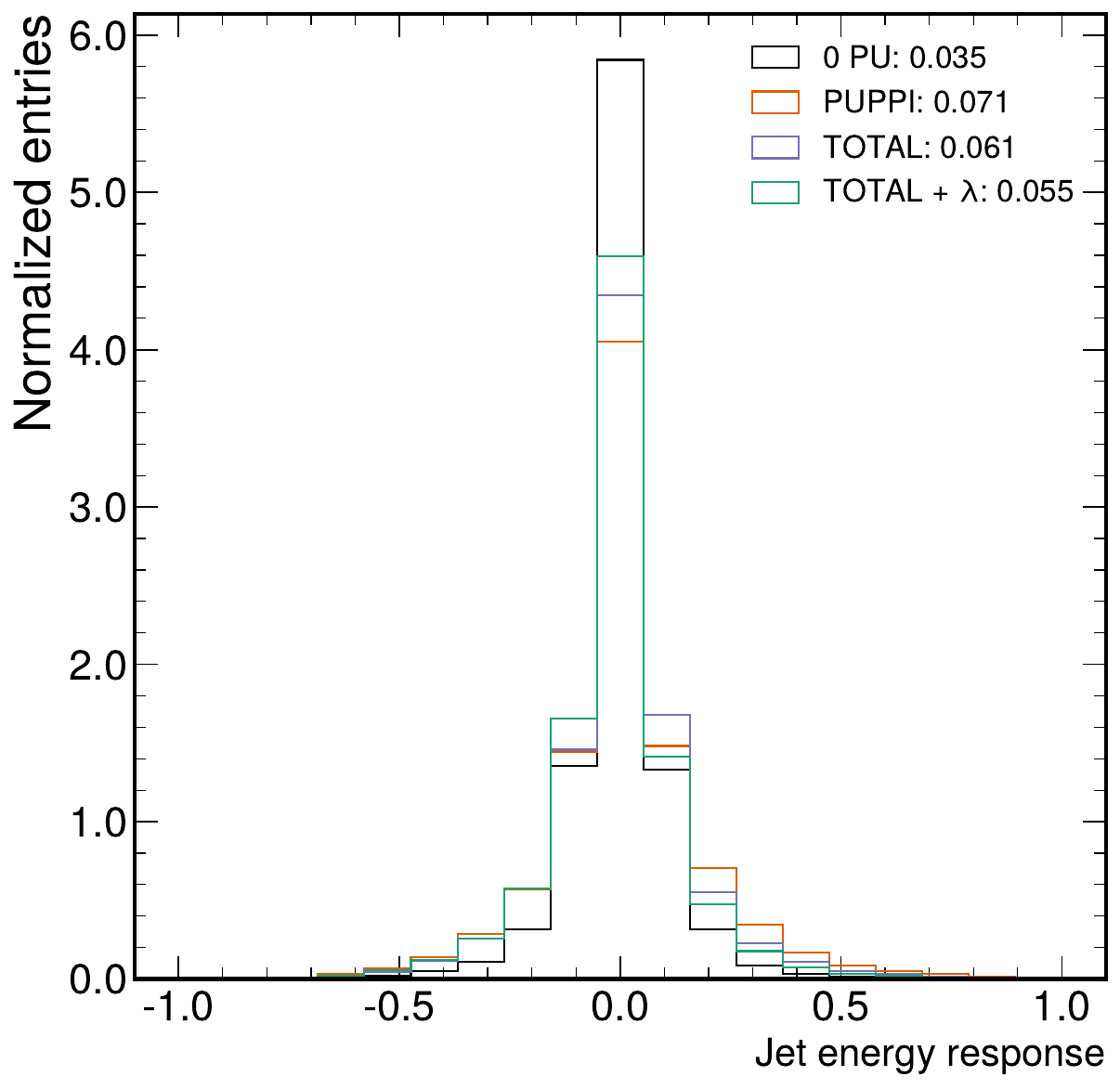}
    \includegraphics[width=0.32\textwidth]{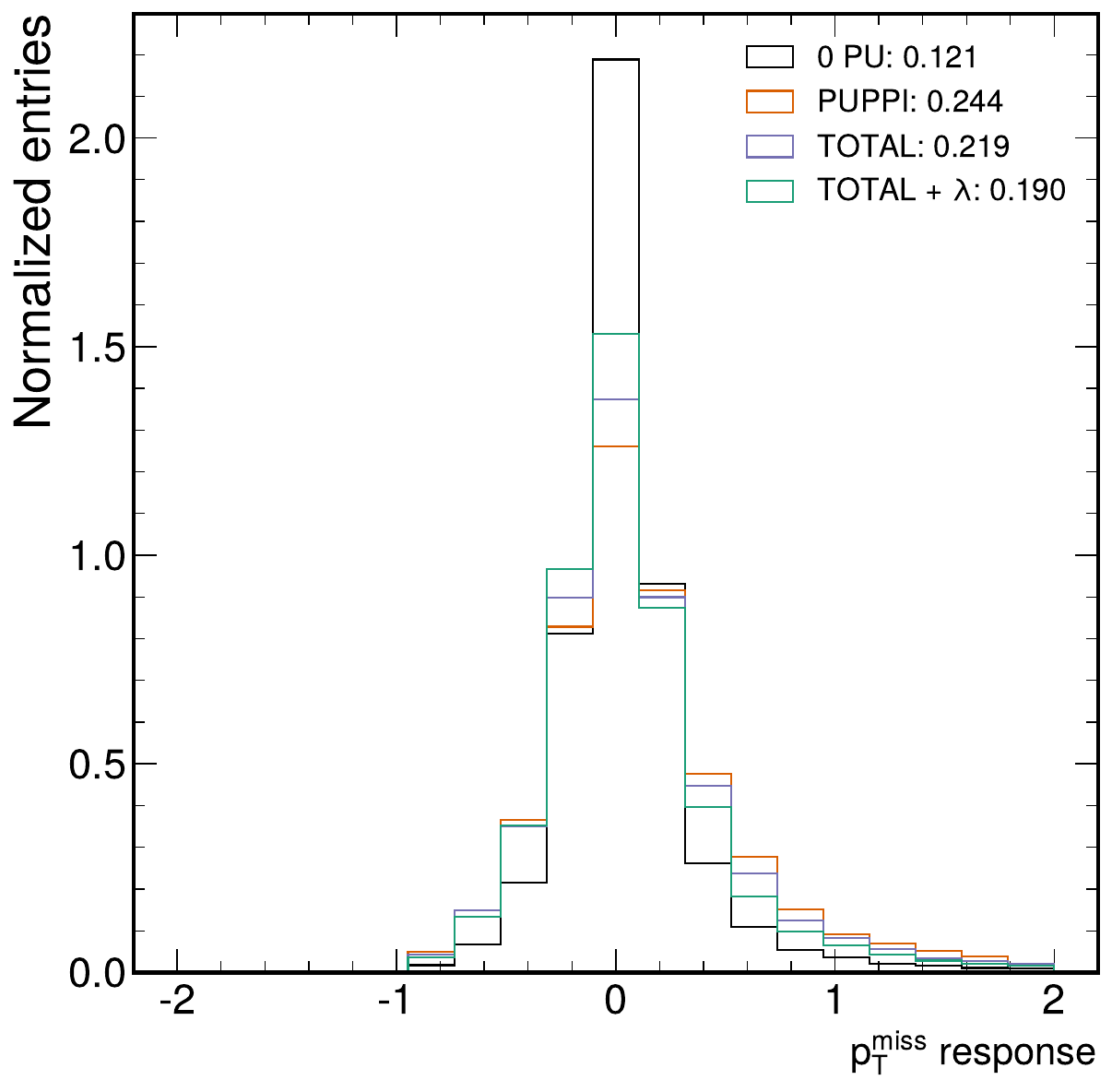}
    \includegraphics[width=0.33\textwidth]{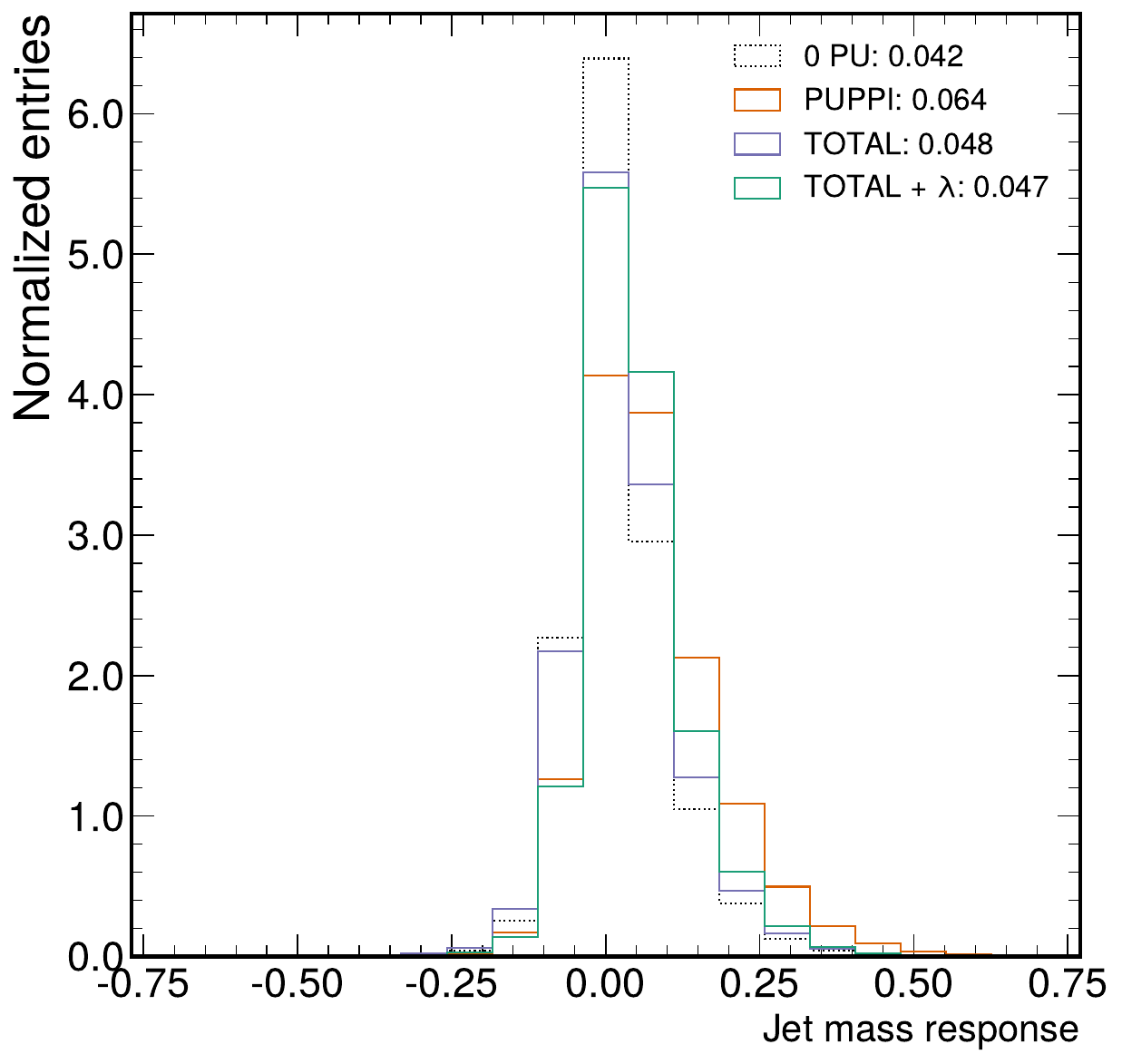}
    \caption{Distributions of the jet \pt and missing transverse momentum \ptmiss for \ttbar events and of the jet mass for $\textrm{Z}^\prime$ events (top row);  and the respective response functions (bottom row). Numbers shown in the response function represent the resolution of the observable.
    We compare the \textsc{TOTAL} algorithm with (green) and without (violet) additional energy conservation, and the \textsc{PUPPI} algorithm (orange), with the ideal scenario of zero pileup interactions (black), and the distribution at generator level (gray). The ratio panels display the percentage difference in resolution of different algorithms compared to generator level events.}
    \label{fig:OTincresolutions}
\end{figure*}

We define the response for a given observable $x$ as the difference between the reconstructed and generator-level values, divided by the generator-level value, $(x_{\mathrm{reco}} - x_{\mathrm{gen}})/x_{\mathrm{gen}}$. In Figure~\ref{fig:OTincresolutions} we show the distributions for different kinematic variables and the corresponding response functions in the \ttbar and $\textrm{Z}^\prime$ samples. The spread of the response is indicative of the experimental resolution of the reconstruction algorithms. We define the resolution in an observable as the spread in its response, given by

\begin{equation}
    \frac{q_{75\%} - q_{25\%}}{2},
\label{eq:resolution}
\end{equation}
where $q_{n\%}$ represents the $n$-th percentile of the response distribution. Lower resolutions result in a better reconstruction and increased power of kinematic variables such as invariant masses or \ptmiss.

\begin{figure*}[htb!]
    \centering
    \includegraphics[width=0.32\textwidth]{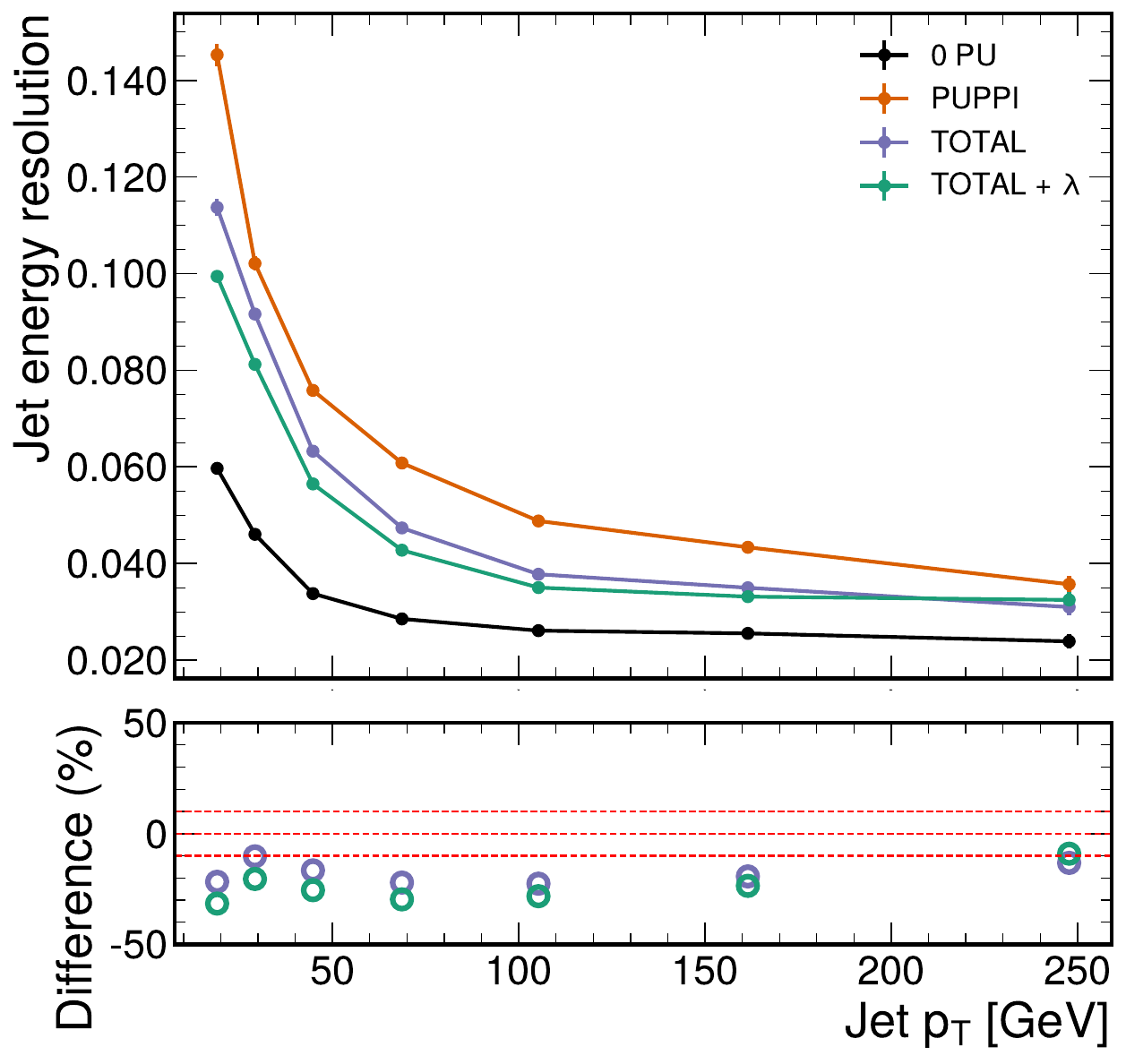}
    \includegraphics[width=0.32\textwidth]{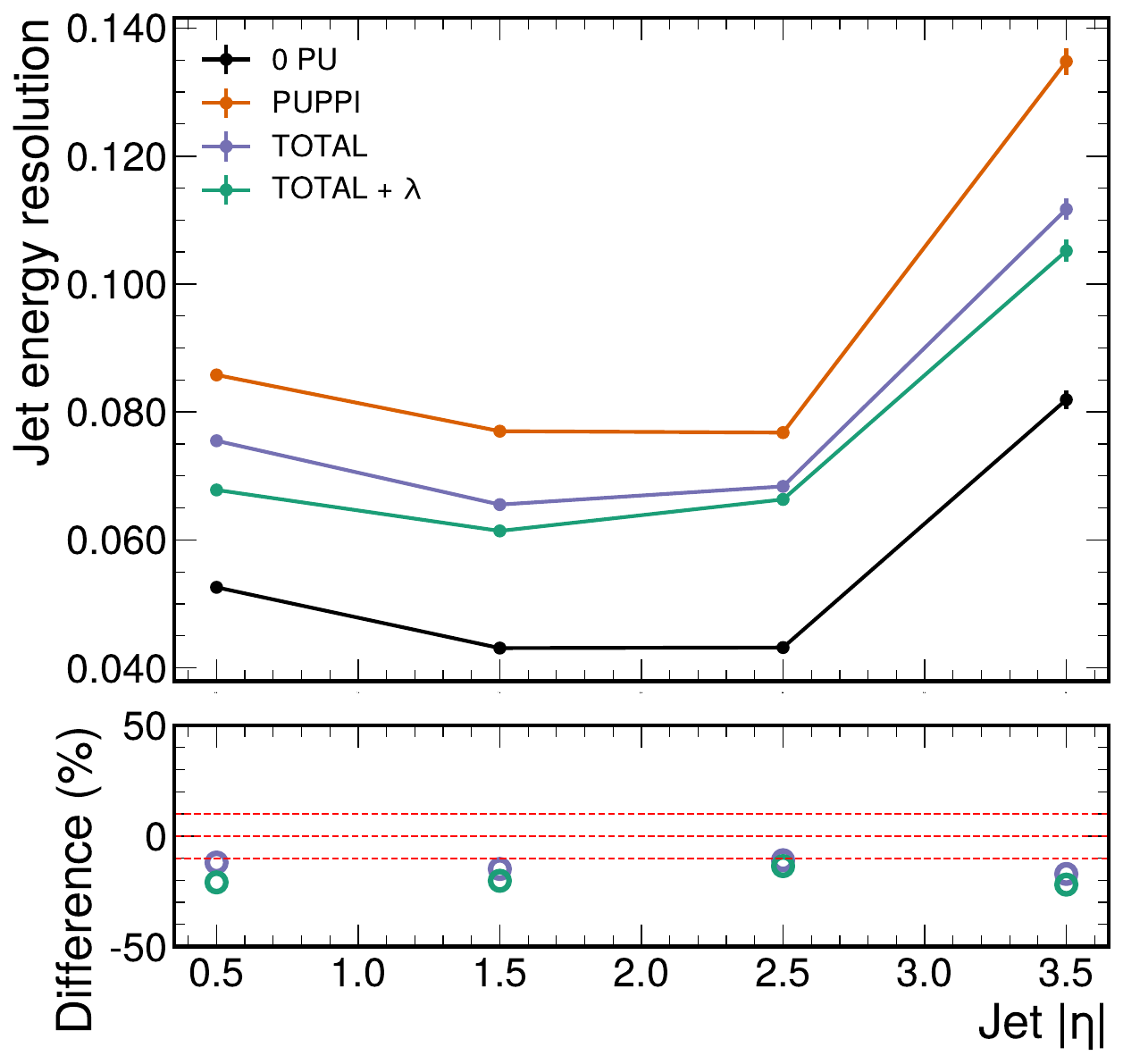}
    \includegraphics[width=0.323\textwidth]{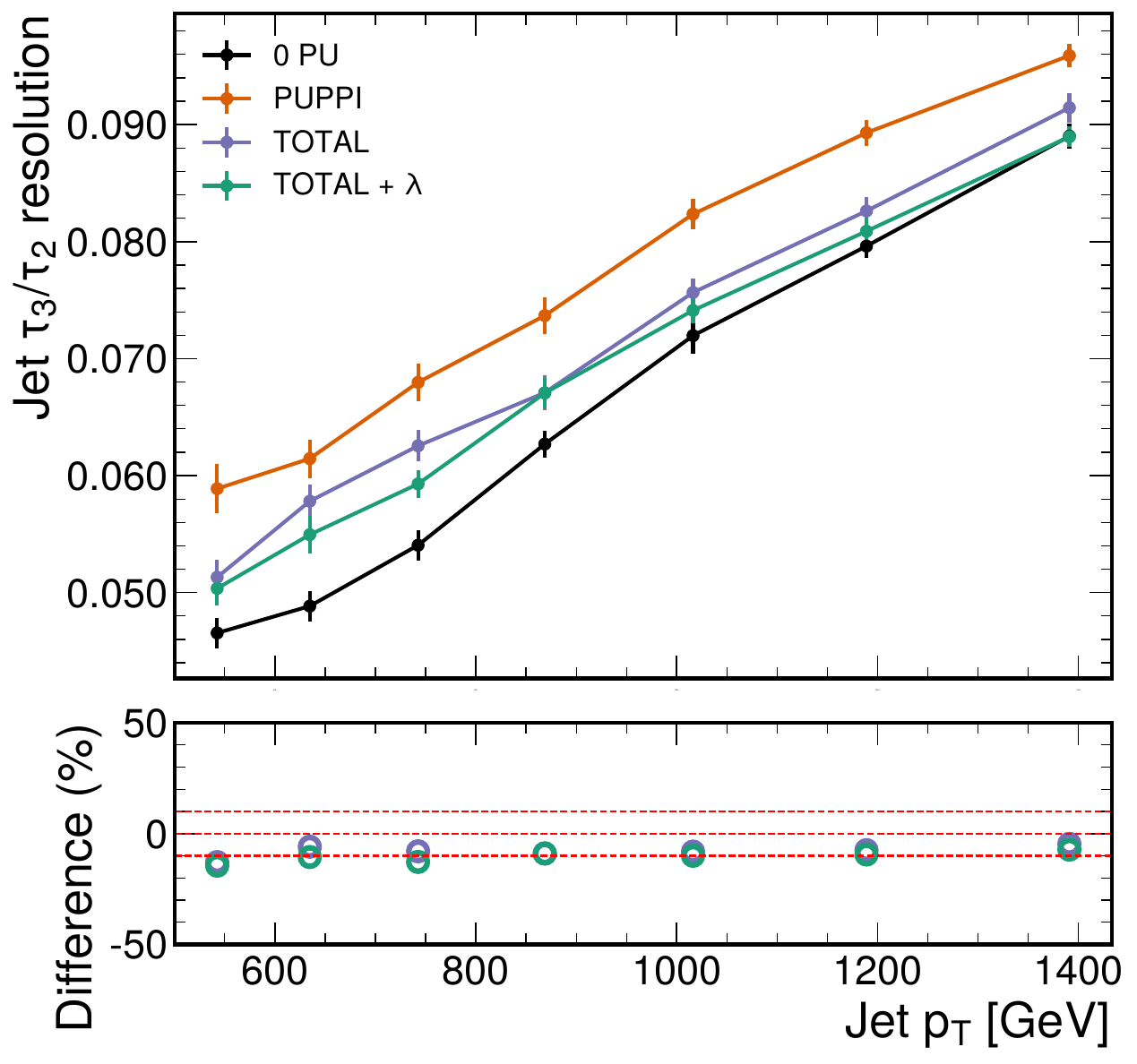}
    \caption{JER as a function of the generator-level jet \pt\ for \ttbar events (left); JER as a function of the generator-level jet $\eta$ for QCD events (center); and large-radius jet $\tau_3/\tau_2$ resolution as a function of the generator-level jet \pt\ for $\textrm{Z}^\prime\to\mathrm{t}\bar{\mathrm{t}}$ events (right). We compare the \textsc{TOTAL} algorithm with (green) and without (violet) additional energy conservation, and the \textsc{PUPPI} algorithm (orange) with the ideal scenario of zero pileup interactions (black). The ratio panels display the percentage difference in resolution of TOTAL compared to PUPPI.}
    \label{fig:OTresolutions}
\end{figure*}

Figure \ref{fig:OTresolutions} shows the jet resolution in the transverse momentum \pt (jet energy resolution, JER) of generator-matched jets, computed as a function of \pt and $\eta$ of the matched generator-level jet, for \ttbar and QCD multijet events, respectively; and the large-radius jet resolution in $\tau_3/\tau_2$, which is a variable that provides a handle on the substructure of the jet~\cite{Thaler:2010tr}, as a function of the \pt of the matched generator-level jet for $\textrm{Z}^\prime$ events. While truth labels for pileup particles are not always available, we investigate the differences in performance of the \textsc{TOTAL} algorithm with respect to a fully supervised network trained using the same strategy as the one used in \cite{Maier:2021ymx}. Results are presented in the Appendix.~\ref{app:supervised}



Across the entire \pt spectrum considered, \textsc{TOTAL} jets are found to have a better JER than \textsc{PUPPI} jets, with improvements of $\sim20\%$ at low \pt. The PUPPI parameters used for the comparison have been taken from~\cite{CMS-PAS-JME-14-001}.  We observe the convergence of all algorithms in the high-\pt regime, as expected due to the reduced effect of soft pileup particles in highly energetic jets. Similarly, we observe a better JER across all the considered $\eta$ spectrum, with improvements of the order of $15\%$ in the forward region of the detector ($|\eta| > 2.5$). The JER degrades in the forward region for all algorithms, due to a worse intrinsic performance of the instrumentation and higher levels of pileup. Finally, we observe a better performance in the $\tau_3/\tau_2$ variable with an improvement up to roughly $10\%$. The degradation of the $\tau_3/\tau_2$ resolution with increasing \pt\ is interpreted as the effect of particles becoming more collimated, resulting in a more challenging definition of the centers of the energy prongs. Finally, we observe that enforcing the energy conservation further improves the performance of \textsc{TOTAL} algorithm in all observables by up to $10\%$.


\begin{figure}[t]
    \centering
    \includegraphics[width=0.45\textwidth]{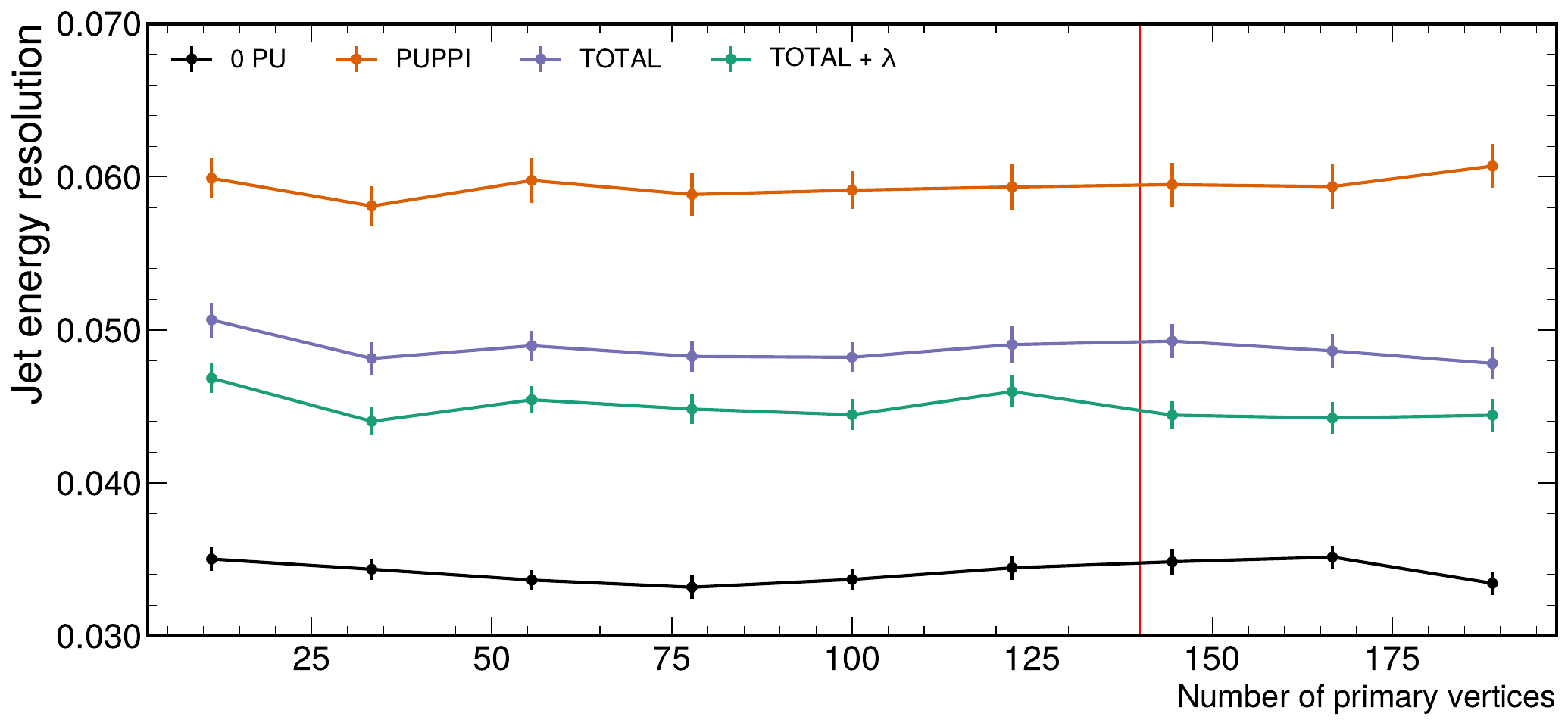}
   
    \caption{Jet energy resolution as a function of the number of pileup interactions for W$+$jets events. We compare the \textsc{TOTAL} algorithm with (green) and without (violet) additional energy conservation, and the \textsc{PUPPI} algorithm (orange) with the ideal scenario of zero pileup interactions (black). The vertical line indicates the average number of PU collisions in the training.}
    \label{fig:vspu}
\end{figure}

A robust pileup mitigation algorithm is required to show stable performance with respect to various physics processes and to the number of pileup interactions in a wide range of values, since pileup conditions can change during data-taking. To check the stability of our algorithm, we evaluate our model on W$+$jets events generated with a uniform distribution in the number of primary vertices, NPV, ranging from 0 to 200, and compute the JER as a function of the number of pileup interactions. In Fig. \ref{fig:vspu}, the performance of \textsc{TOTAL} is found to be stable and consistently better than \textsc{PUPPI} across the entire NPV spectrum. This testifies to the ability of the network to adapt to pileup scenarios and processes different from the ones experienced during training. 
 
\begin{figure}[t!]
    \centering
    \includegraphics[width=0.475\textwidth]{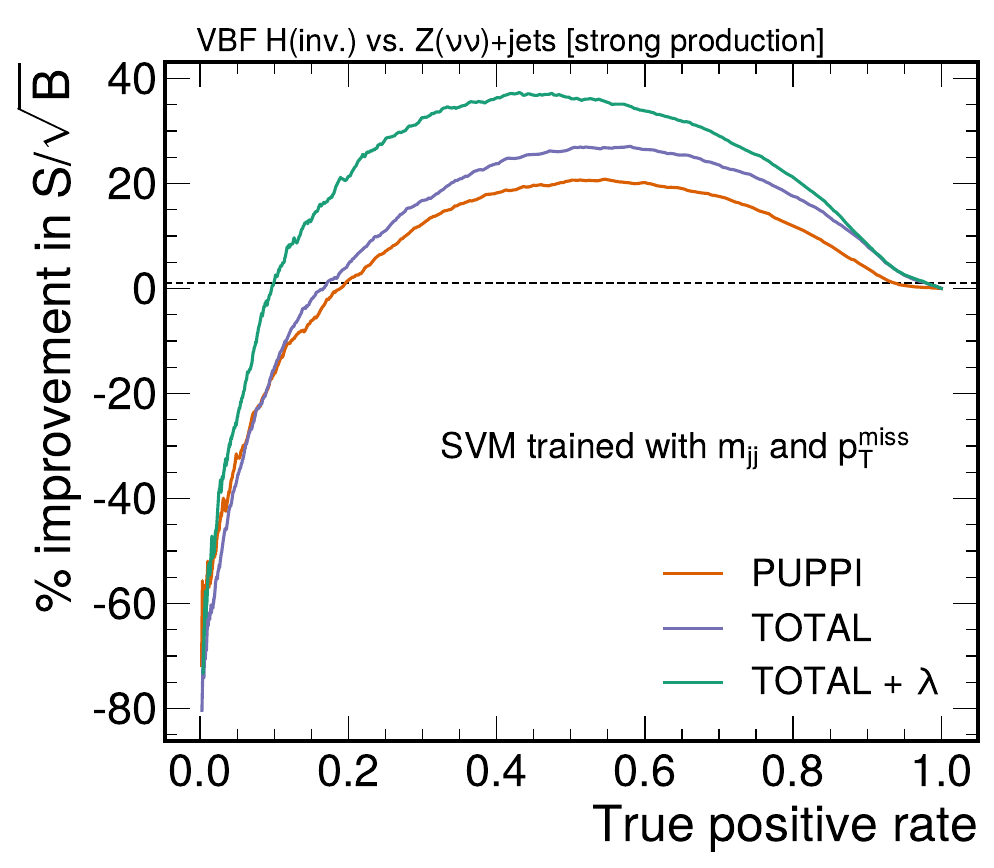}\\
     \caption{A linear classifier (Support Vector Machine, SVM) trained with \ptmiss and the dijet mass results in a sizable improvement in $S/\sqrt{B}$ for the TOTAL algorithms over PUPPI.}
    \label{fig:MET_plus_OTsic}
\end{figure}

Finally, to illustrate the benefit TOTAL brings to searches for new physics, we study its impact in a search for invisible Higgs boson decays, which is one of the essential search channels at the LHC.  Such decays are highly suppressed in the Standard Model, rendering any observation of an enhanced decay rate an unambiguous sign for new physics. Here, the Higgs boson is assumed to be produced via the fusion of two vector bosons in association with two jets close to the beam axis. A predominant background in this search is the production (mediated by the strong interaction) of a Z boson decaying into neutrinos. In Fig.~\ref{fig:MET_plus_OTsic}, we demonstrate the improvement in significance $S/\sqrt{B}$, with $S$ being the signal yield and $B$ the background yield, as a function of a selection on a linear classifier constructed from \ptmiss and the dijet mass. We find the improvement from TOTAL to be in the order of 15\% compared to PUPPI, consistent with the improvements in resolution for jets and \ptmiss. This would lead to a better sensitivity in the search for such decays and significantly improve the expected upper limit on the branching ratio of Higgs bosons to invisible particles quoted by the ATLAS and CMS collaborations~\cite{atlasvbfinv2022,PhysRevD.105.092007}.

\section{Conclusions}
\label{sec:conclusions}

We have presented a novel method to identify particles from the primary interaction and reject pileup particles in proton collisions at hadron colliders. Without relying on training labels, our self-supervised algorithm is able to morph an inclusive particle collection into a collection containing just the products of the primary interaction. Our algorithm provides a perspective for mitigating the impact of pileup at the High-Luminosity LHC, yielding an improvement over the current state-of-the-art of up to 25\% in the resolution of key observables used in searches for new physics and in precision measurements. To the best of our knowledge, this is the first application of optimal transport concepts embedded into graph neural networks to solve a highly relevant problem of current and future particle collider experiments. Not relying on per-particle truth labels, our approach can be implemented in full-scale simulations of detectors such as ATLAS and CMS. Thus, we encourage our colleagues to test this strategy. Additionally, our approach can be applied to other problems inside and outside the field of high energy physics where de-noising is of great importance and where a realistic simulation of noise exists even if the noise distribution is intractable. For instance, it could be used for shower reconstruction in highly granular calorimeters, or for mitigating noise in time series-based astronomical data. Future studies should focus on the expansion of the method to compare statistically independent samples with different noise levels. This would allow for a data-driven, global determination of pileup at the LHC, e.g., by comparing distributions obtained at the beginning of a given proton fill and towards the end, when less pileup is present in the data.

\section{Acknowledgments}
FI is supported by the Ministry of Science and Technology of China, project No. 2018YFA0403901 and National Natural Science Foundation of China, project No. 12188102, 12061141003. This research used resources of the National Energy Research Scientific Computing Center, a DOE Office of Science User Facility supported by the Office of Science of the U.S. Department of Energy under Contract No. DE-AC02-05CH11231 using NERSC award HEP-ERCAP0021099. This research was supported in part by the Swiss National Science Foundation (SNF) under project No. 200020\_204975/1. This project was supported by funding from the Alexander von Humboldt Foundation.
\nocite{*}

\bibliography{references}

\appendix

\section{Appendix. Comparison with a fully supervised network}\label{app:supervised}

\begin{figure*}
    \centering
    \includegraphics[width=0.475\textwidth]{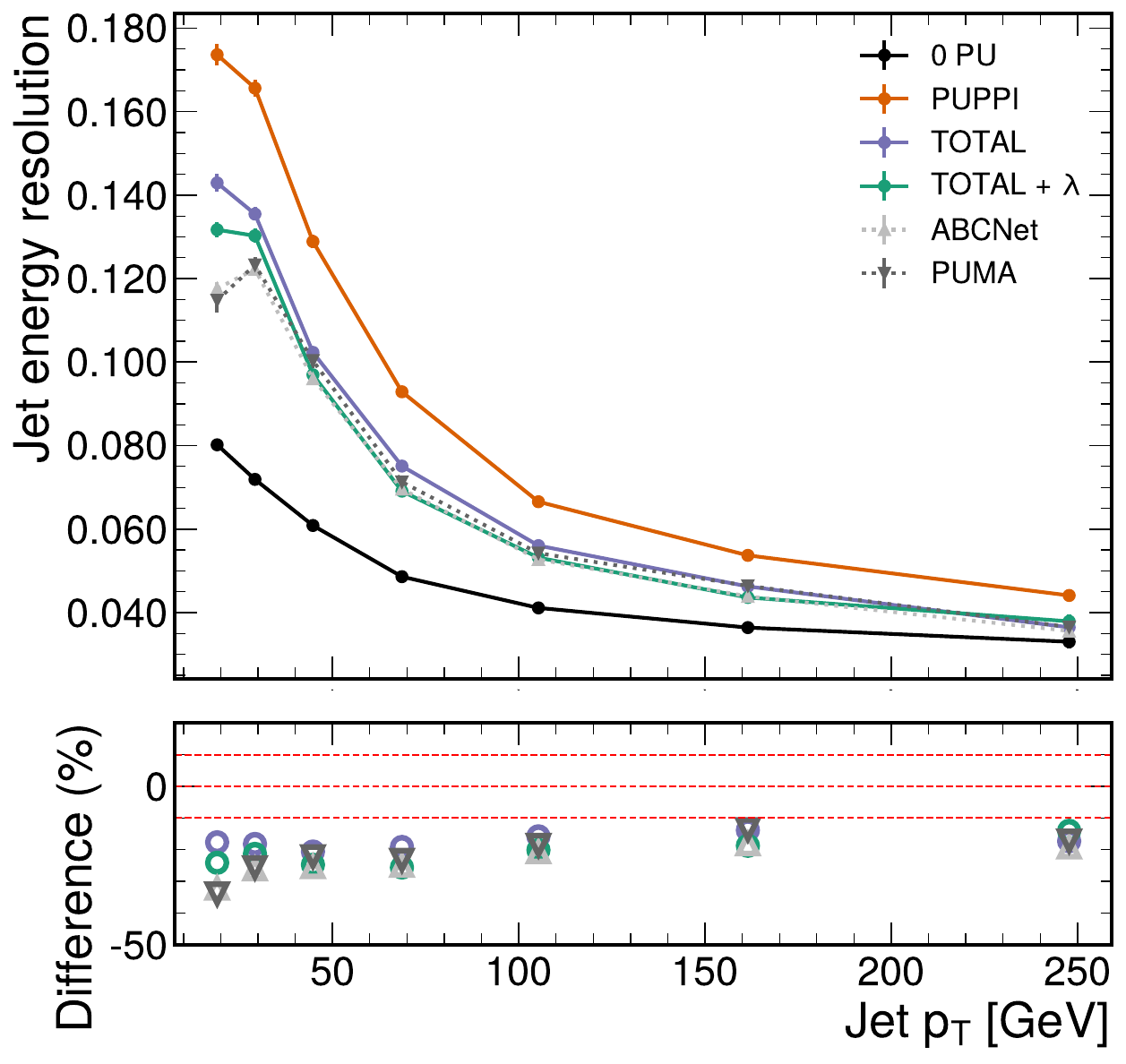}
    \includegraphics[width=0.46\textwidth]{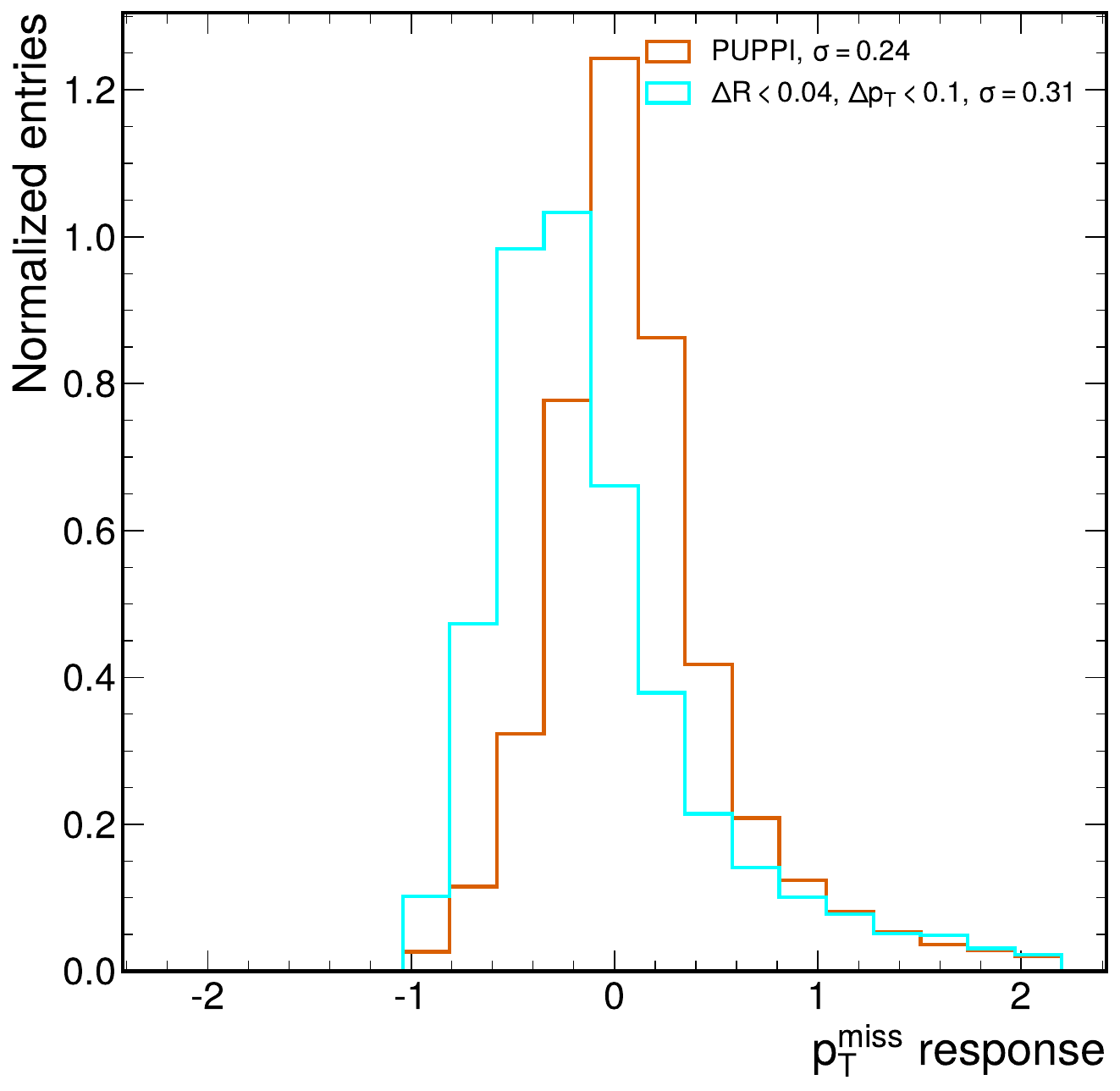}\\
     \caption{Left: jet energy resolution for QCD events as a function of the generator level \pt. We compare the \textsc{TOTAL} algorithm with (green) and without (violet) additional energy conservation, and the \textsc{PUPPI} algorithm (orange) with the ideal scenario of zero pileup interactions (black). Additional supervised results are shown for \textsc{ABCNet} (light gray) and \textsc{PUMA} (dark gray), both using unrealistic, perfect labels. Right: comparison of \ptmiss resolutions for PUPPI and for a human-annotation-based per-particle truth, where particles are labeled as coming from the primary interaction if they can be matched to a particle in the same event without pileup overlaid, and from pileup otherwise. The poor resolution when using truth definition renders fully supervised algorithms relying on these targets inferior to our self-supervised strategy.}
    \label{fig:dijet_sup}
\end{figure*}
The \textsc{TOTAL} algorithm is trained using a self-supervised setting to avoid the need of labels from human annotation that cannot perfectly be obtained in full simulation or in data. However, to compare our results with a fully supervised setting using perfect labels, which are available in the simplified reconstruction of the Delphes framework, we retrain the backbone architecture of \textsc{TOTAL}, \textsc{ABCNet}, using the same strategy presented in~\cite{Maier:2021ymx}, with a regression objective that aims to learn the energy fraction of the primary collision carried by each individual particle and resulting in 0 labels for pileup interactions. Similarly, we also compare with the \textsc{PUMA} algorithm to check the differences in performance obtained by the use of different backbone architectures. Results for the jet energy resolution versus the jet transverse momentum for QCD events are shown in Figure~\ref{fig:dijet_sup}. We observe the resolution obtained by \textsc{TOTAL} to be similar to the one obtained by the supervised training of both \textsc{ABCNet} and \textsc{PUMA}, with differences in performance below 10\% in all generator level \pt intervals. 
Moreover, Figure~\ref{fig:dijet_sup} demonstrates the performance of a truth definition obtained by matching (in angular separation and momentum difference) particles between the samples with and without pileup contributions to identify particles coming from the hard interaction. For the matching to be successful, particles have to be matched within $\Delta R=0.04$ and have transverse momenta compatible within 10\%. A successful matching of particles could then be used as truth labels for a fully supervised algorithm \`a la PUMA. However, as is visible from a comparison with even the classical benchmark PUPPI, the performance in \ptmiss resolution is severely degraded for such a truth definition even after cumbersome manual tuning of the matching parameters, rendering the application of fully supervised algorithms an arduous if not impossible task if one wants to achieve results competitive with our self-supervised approach.

\end{document}